# A Note on Publicly Verifiable Quantum Money with Low Quantum Computational Resources

Fabrizio Genovese[1], and Lev Stambler[1,2]

[1] NeverLocal LTD, , London, UK
{fabrizio,lev}@neverlocal.com
https://www.neverlocal.com
[2] University of Maryland, College Park,

**Abstract.** In this work we present a publicly verifiable quantum money protocol which assumes close to no quantum computational capabilities. We rely on one-time memories, which in turn can be built from quantum conjugate coding and hardware-based assumptions, together with preimage-resistant hash functions. Specifically, our scheme allows for a limited number of verifications and also allows for quantum tokens for digital signatures. Double spending is prevented by the no-cloning principle of conjugate coding states. An implementation of the concepts presented in this work can be found at https://github.com/neverlocal/otm_billz.

## 1 Introduction

*Quantum money* is not a new concept. Originally proposed by Wiesner around 1970[1] [1], it predates blockchain by at least 35 years. Though the original idea is quite simple, elegant, and has been prototyped [2, 3], it has a key limitation: it is a *private key* quantum money scheme. This means that only the party that mints the money can verify its authenticity and so each use of the money requires contact with the issuer.

Public key quantum money [4], on the other hand, allows anyone to verify the authenticity of the money without needing to contact the issuer. This is a crucial property for a usable currency, but it has proven to be quite elusive to implement.

Quantum money relies on unique properties of quantum resources, primarily their non-clonability, to solve the double spending problem. In layman terms, a quantum money scheme provides:

1. A way to 'mint' monetary value into quantum resources. These quantum resources become akin to banknotes and can be transferred to other parties - for example via *quantum teleportation* [5]; The *no-cloning theorem* [6] guarantees protection against double-spending;

[1] Wiesner's work on quantum money remained unpublished until 1983.

2. A way to verify the authenticity of these quantum banknotes, meaning verifying that they have been issued by a trusted party, or by following a given protocol. Ideally, to minimize trust, a given party should be able to establish the banknotes' authenticity and viability *before* receiving them.

About point 2, notice that in this work we won't be concerned with who actually mints the money; this could be a trusted party (e.g. a central bank or a stablecoin issuer) or a decentralized protocol rewarding contributors (e.g. PoW [7]).

The great upside of quantum money compared to blockchain-based payment schemes is that **it does not require a consensus protocol**. Quantum money works exactly as cash, and as such it is intrinsically private and peer-to-peer, and does not require setting up any system of economic incentives to be run.

Since Wiesner's work, many people have contributed to this line of research, often relying on heavy quantum computation which we do not yet have [4, 8, 9]. In this work, we present a publicly verifiable quantum money scheme that supports up to $N$ verifications, for some parameter $N$. Importantly, **we assume minimal quantum computational capabilities**: parties need only the ability to *prepare*, *transmit*, and *measure* single-qubit states in two conjugate bases. This is precisely the capability provided by *quantum information and communication equipment*, such as the one used for quantum key distribution (QKD), which is orders of magnitude more accessible than a general-purpose *quantum processor (QPU)*.

The practicality of our scheme depends on the availability of quantum internet and on the reliability of quantum memories. The former is growing very fast, with quantum internet services being already planned for data centers [10] and entire countries [11–14]. As for the latter, the capability of holding a quantum state has been steadily growing, with $T_2$ coherence times for some technologies exceeding 10 hours [15], though such times currently require extreme cryogenic conditions. Stable milli-second coherence is already sufficient for high-frequency trading applications. While significant engineering challenges remain, we believe quantum money represents a compelling near-term application for emerging quantum infrastructure.

Technically, we proceed as follows:

- In Section 2 we provide a gentle introduction to quantum information theory and to *conjugate coding*;
- In Section 3 we recap the idea of building *One-Time Memories (OTMs)* using conjugate coding and *secure hardware*. Some sort of hardware assumption is necessary since OTMs are provably impossible both in the classic and ideal quantum setting [16, 17].
- In Section 4 we propose a public key quantum money scheme.
- In Section 5 we show how our scheme naturally generalizes to *quantum tokens* [18], and provide applications;
- Finally, in Section 6 we expand on the current bottlenecks to make our scheme practical, and on directions of future work.

As we deem this a short applications paper almost all the relevant proofs will be referenced in the bibliography, with the exception of some original proof sketches.

As we remarked already in the abstract, we implemented the concepts hereby presented into a proof-of-concept rust library, which can be found at https://github.com/neverlocal/otm_billz.

## 2 Quantum Recap

Quantum and classical information differ fundamentally. Classical information can be freely copied, accessed and acted upon; quantum information cannot. The *no-cloning theorem* states that arbitrary quantum information cannot be copied; furthermore, accessing quantum information, known as *measurement*, converts it to classical information and alters the original quantum information forever.

We will follow the so-called *Dirac-Von Neumann picture* [19, 20] for a formal introduction of these concepts. First and foremost, each quantum system is modelled as a *separable complex Hilbert space* $H$, and a *state* of such system $\psi$ is represented as a *normalized vector* in $H$, up to scalar multiples. We denote a normalized vector as $|\varphi\rangle$, and its dual as $\langle\varphi|$. The *inner product* $\langle\psi^*, \varphi\rangle$ is denoted $\langle\varphi|\psi\rangle$, whereas $|\psi\rangle\langle\varphi|$ stands for the outer product.

In quantum information theory we usually operate on *qubits*, which are two-dimensional Hilbert spaces $\mathbb{C}^2$. Normalized vectors in $\mathbb{C}^2$ correspond to the points of the *complex projective line* $\mathbb{C}\boldsymbol{P}^1$, which can be *embedded* in 3D space allowing us to represent a qubit state $|\psi\rangle$ as a point on a sphere, known as the *Bloch sphere* (see Fig. 1a). It is commonplace to also display the following states on the sphere: $|0\rangle$ and $|1\rangle$ at the north and south poles respectively, which are the usual computational basis $\binom{1}{0}, \binom{0}{1}$ on $\mathbb{C}^2$ and are known as the *Z-basis*; $|+\rangle, |-\rangle$ at the west and east poles respectively, which are defined as $\frac{|0\rangle+|1\rangle}{\sqrt{2}}$ and $\frac{|0\rangle-|1\rangle}{\sqrt{2}}$ and known as the *X-basis*.

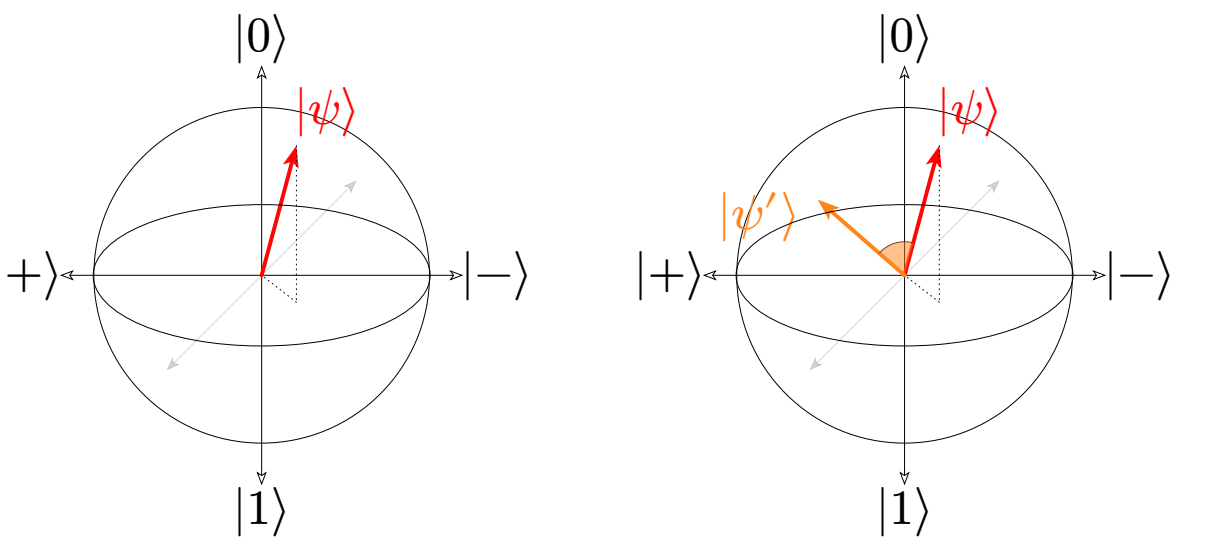

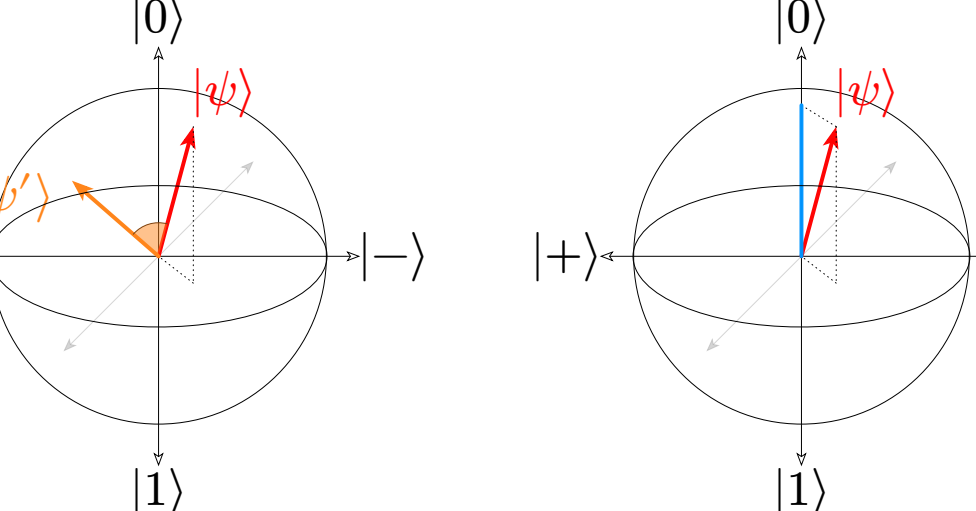

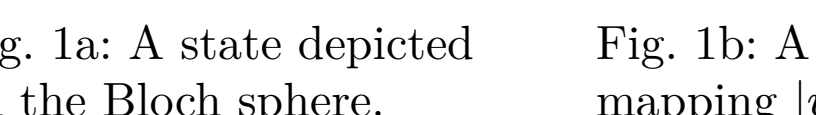

Fig. 1a: A state depicted on the Bloch sphere.

Fig. 1b: A transformation mapping $|\psi\rangle \mapsto |\psi'\rangle$.

Fig. 1c: A quantum measurement in the Z-basis.

A system consisting of $N$ qubits is taken to be the *n-th tensor product* of $\mathbb{C}^2$, namely $\otimes_N \mathbb{C}^2$. When operating on multiple qubits, a state $|\psi\rangle$ will often implicitly be taken to mean $|\psi\rangle_1 \otimes ... \otimes |\psi\rangle_N$ for some $N$. Notably, the dimension of a $N$

qubit system is $2^N$, that is, the (maximum) number of orthogonal states a $N$ qubit system can be in **grows exponentially** with $N$.

*Transformations* on a quantum system $H$ are taken to be *unitaries*; in the case of a single qubit, these are the elements of the group $SU(2)$. The isomorphism $SU(2)/\{\pm 1\} \simeq SO(3)$ guarantees that every admissible transformation on a qubit uniquely corresponds to some rotation of the Bloch Sphere (depicted on Fig. 1b).

In the context of quantum physics, *measuring an observable* leads to probabilistic results. This means that, for instance, if we measure some property of a given system we will obtain some outcomes with some probability. On paper, what we can do is *computing the expected value of an observable*, that is, compute said probabilities. This entails choosing some *self-adjoint operator* $A$ on our space (representing the "quantity" we want to measure) and, given a normalized state $\psi$ (the state we want to measure), computing the inner product $\langle \psi^*, A\psi \rangle$, denoted $\langle \psi | A | \psi \rangle$. In the case of qubits, this can be simplified: Given a state $|\psi\rangle$ and choosing an orthonormal basis (e.g. the Z-basis $|0\rangle, |1\rangle$), we can write $|\psi\rangle = \alpha|0\rangle + \beta|1\rangle$. $|\psi\rangle$ is normalized, so $\alpha^*\alpha + \beta^*\beta = 1$. We take $\alpha^*\alpha$ and $\beta^*\beta$ to represent the *probabilities* that $|\psi\rangle$ yields outcome 0 or outcome 1, respectively, when measured along the Z-basis. After measurement, $|\psi\rangle$ collapses[2] to either $|0\rangle$ with probability $\alpha^*\alpha$; or to $|1\rangle$ with probability $\beta^*\beta$. This means that the original state $|\psi\rangle$ **ceases to exist**, and irreversibly becomes one of the possible measurement outcome states with some probability.

On the Bloch sphere, we obtain the probability for a state $|\psi\rangle$ to yield a given measurement outcome by taking the axis spanned by a given basis (orthonormal basis vectors are always antipodal on the Bloch sphere) and projecting the vector over that axis (shown in Fig. 1c).

### 2.1 Conjugate Coding

The basis vectors $|+\rangle, |-\rangle$ are *maximally unbiased* with respect to the basis vectors $|0\rangle, |1\rangle$: Measuring either $|+\rangle$ or $|-\rangle$ along the Z-basis yields $|0\rangle$ or $|1\rangle$ with probability $\frac{1}{2}$; Similarly, $|0\rangle, |1\rangle$ can be expressed in the X-basis as $\frac{|+\rangle}{\sqrt{2}} + \frac{|-\rangle}{\sqrt{2}}$ and $\frac{|+\rangle}{\sqrt{2}} - \frac{|-\rangle}{\sqrt{2}}$ respectively, yielding again probabilities of $\frac{1}{2}$ when measured in the X-basis. In the Bloch sphere, this is evident as the axes spanned by the Z- and X-basis are orthogonal. This informs the following construction:

**Construction 2.1.1.** *Given a uniformly chosen bit $\theta$, map a bit $b \in \{0, 1\}$ to a qubit:*
- *If $\theta = 0$: If $b = 0$ output $|b\rangle_0 := |0\rangle$; if $b = 1$ output $|b\rangle_0 := |1\rangle$;*
- *If $\theta = 1$: If $b = 0$ output $|b\rangle_1 := |+\rangle$; if $b = 1$ output $|b\rangle_1 := |-\rangle$.*

Notice that given $|b\rangle_\theta$ it is impossible to recover $b$ without knowing $\theta$: Measuring a qubit in the right basis entails recovering the bit with certainty (e.g. measuring in the Z-basis if $\theta = 0$), whereas choosing the wrong basis (e.g. the Z-basis if $\theta = 1$) outputs either 0 or 1 with probability $\frac{1}{2}$, which is just noise. We can extend

[2] A formally satisfying treatment of measurement collapse requires *density matrices*, *POVMs* and *Kraus operators*. We redirect the interested reader to [21].

this construction to transmit two classical bits in such a way that only one can be recovered.

**Construction 2.1.2.** *(Conjugate coding [1].) Given a uniformly chosen bit $\theta$, map two bits $b, b'$ to two qubits by:*

- *If $\theta = 0$:*
  - *Map $b$ to $|0\rangle$ if it is $b = 0$, to $|1\rangle$ if it is $b = 1$. This is the first qubit;*
  - *Map $b'$ to $|+\rangle$ if it is $b' = 0$, to $|-\rangle$ if it $b' = 1$. This is the second qubit;*
- *If $\theta = 1$, swap the order of the two qubits above.*

Again, without knowing $\theta$, any receiving party can extract the classical data accurately only $\frac{3}{4}$ of the times [1]. If the receiving party gets $\theta$ **after** performing a measurement, we can be sure that only one bit will be recovered. If, say, the receiving party wants to recover $b$, it will measure both qubits in the Z-basis. $\theta$ will provide information about which one of the outcome measurements is noise. This idea is at the basis of the work on *one time memories* presented in [22].

## 3 One-Time Memories from Conjugate Coding

A *one-time memory (OTM)* is a cryptographic primitive that allows a sender to encode two secret values $s^0, s^1$ into a device such that a receiver can retrieve exactly one of them, but not both. In a line of works [23–25], Liu introduced the idea of constructing OTMs from conjugate coding combined with hardware assumptions.

A series of work follow up on Liu's idea, improving the security and also relying on various hardware assumptions, including but not limited to: (1) secure, stateless hardware [18, 22, 26], (2) noisy quantum storage [27], (3) limited quantum computational power [28, 29].

To highlight the main idea and construction simplicity of OTMs based on conjugate coding, we briefly summarize the one-time memory construction of Ref. [22] and Ref. [18] which assume stateless and secure hardware. By *secure hardware*, we mean tamper-resistant devices that correctly execute their programmed functionality without leaking internal state. Practical instantiations include *Trusted Execution Environments (TEEs)* such as Intel SGX or ARM TrustZone. While TEEs have known side-channel vulnerabilities [30], mitigations exist, and more importantly, the security of our scheme degrades gracefully: compromising individual hardware tokens affects only the banknotes they protect. We note that **any simulation-secure OTM construction** (see Appendix A) **can be used** in our quantum money scheme.

**Construction 3.3.** *(Sketch of Conjugate-coding OTMs) For bitstrings $s^0, s^1$ from the environment and security parameter $\zeta$: select random $b, \theta \in \{0, 1\}^\zeta$ and proceed as follows:*

- *Hard-code* $b, \theta, s^0, s^1$ *into the secure hardware*[3] *;*
- *Repeat Construction 2.1.2 over the security parameter to build states* $|b_i\rangle_{\theta_i}$*, which are sent to the receiving party along with the hardware token.*
- *The receiving party measures the states* $|b_i\rangle_{\theta_i}$ *in either the Z- or the X- basis, depending on which bitstring,* $s^0, s^1$*, must be retrieved. Roughly half of these measurements will be random noise, while the other half will correspond to the actual bits in* $b$*.*
- *The hardware token then verifies that the non-noise portion of the measurement outcomes corresponds to the respective bits in the string* $b$*, and outputs the corresponding* $s^i$ *if the verification passes.*

As shown in [18], the above construction is secure assuming the security of the hardware and a poly-time adversary. Moreover, the construction is *noise tolerant*, in the sense that it can be proved secure even if the quantum channel between sender and receiver is noisy, up to a certain threshold.

## 4 Public Key Quantum Money

In this section, we show how one-time memories and a preimage-resistant hash function, which in particular implies a one-way function and hence post-quantum secure digital signatures [31], are all we need to build quantum money with a limited number of verifications and minimal quantum computational resources.

At a high level, our scheme combines *cut-and-choose* methods from maliciously secure MPC [32–34] with a MAC-like authentication mechanism where hash pre-images serve as unforgeable tags [35]. The cut-and-choose idea is simple: rather than checking every component (which would consume the entire banknote), we randomly sample a subset to verify. If any fraud exists, it will be detected with high probability. Each one-time memory encodes a pair of hash pre-images, and the banknote consists of multiple such OTMs along with the corresponding hashes (signed by the mint). To verify, a party randomly selects a subset of OTMs to "challenge" by opening them and checking that the revealed pre-images hash correctly. The hash pre-images function like MACs: computationally hard to forge without the OTM, yet easy to verify. Unopened OTMs are retained for future verifications.

[3] In the real world, trusted hardware, such as TEEs, can receive an encryption of $b, \theta, s^0, s^1$ and decrypt them internally.

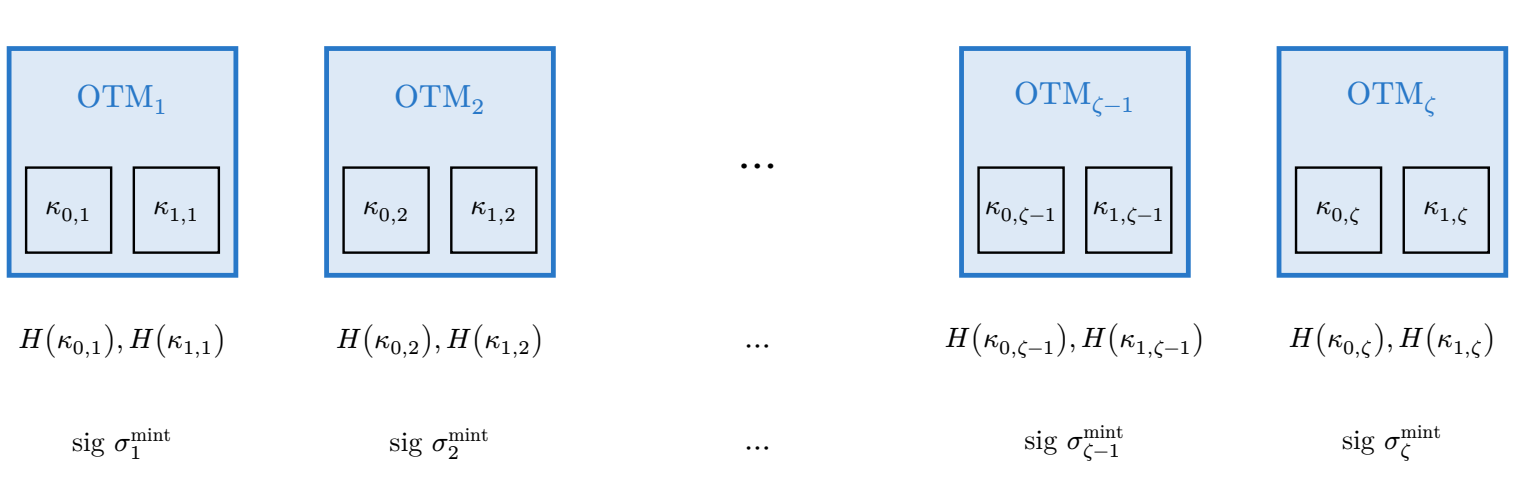

**Fig. 2.** Structure of a quantum banknote with $\zeta$ OTMs. Each $\mathtt{OTM}_j$ encodes two hash pre-images $\kappa_{0,j}, \kappa_{1,j}$. Blue indicates quantum components; black indicates classical data.

**Construction 4.4.** *(Minting Quantum Money)* *Given a* preimage-resistant hash function $H$ *and a security parameter* $\zeta$*, the* mint[4] *generates a* quantum banknote *by:*

- *Uniformly choosing bitstrings* $\kappa_{0,i}, \kappa_{1,i}$ *for* $i \in [\zeta]$*. We require that* $|\kappa_{b,i}| - |H(\kappa_{b,i})| = \omega(\log \lambda)$ *for* $b \in \{0,1\}$ *and each* $i$*, ensuring that pre-images have superlogarithmic min-entropy conditioned on their hashes;*
- *For each* $\kappa_{b,i}$*, computing* $H(\kappa_{b,i})$ *and signing the hash to obtain signature* $\sigma_{b,i}^{\mathtt{mint}}$*;*
- *For each couple* $(\kappa_{0,i}, \kappa_{1,i})$*, generating a one-time memory* $\mathtt{OTM}_i$ *together with its states* $\{|\mathtt{OTM}\rangle_i\}_i$ *as in Construction 3.3, such that* $\mathtt{OTM}_i(b) = \kappa_{b,i}$*.*
- *The tuple* $\left(\left\{H(\kappa_{b,i}), \sigma_{b,i}^{\mathtt{mint}}\right\}_{b,i}, [\zeta], \{\mathtt{OTM}_i, |\mathtt{OTM}\rangle_i\}_i, \zeta, \{\}\right)$ *is the* fresh *quantum banknote.*

In general, a banknote has the form

$$\left(\left\{H(\kappa_{b,i}), \sigma_{b,i}^{\mathtt{mint}}\right\}_{b\in\{0,1\}, i\in[\zeta]}, \mathcal{J}, \left\{\mathtt{OTM}_j, |\mathtt{OTM}\rangle_j\right\}_{j\in\mathcal{J}}, \zeta, \{\kappa_k\}_{k\in\mathcal{K}}\right),$$

where $\mathcal{J}, \mathcal{K} \subseteq [\zeta]$ are such that $\mathcal{J}, \mathcal{K}$ partition $[\zeta]$, i.e., $\mathcal{K} = [\zeta] \setminus \mathcal{J}$.

Any party can verify that a given quantum banknote is original as follows.

**Construction 4.5.** *(Verifying Quantum Money)* *Given a party* $V$*, a cut-and-choose parameter* $\xi < \zeta$*, and a quantum banknote* $\left(\left\{H(\kappa_{b,i}), \sigma_{b,i}^{\mathtt{mint}}\right\}_{b,i}, \mathcal{J}, \left\{\mathtt{OTM}_j, |\mathtt{OTM}\rangle_j\right\}_j, \zeta, \{\kappa_k\}_k\right)$

[4] Again, this could be a trusted party (a bank) or the result of some MPC effort.

- *$V$ checks* well-formedness*: that $\mathcal{J} \cup \mathcal{K} = [\zeta]$ and $\mathcal{J} \cap \mathcal{K} = \emptyset$ (i.e. $\mathcal{J}, \mathcal{K}$ form a partition of $[\zeta]$), that the number of OTM instances equals $|\mathcal{J}|$, and that the number of revealed pre-images equals $|\mathcal{K}|$ with each pre-image mapping to one of the signed hashes;*
- *$V$ verifies that the signatures $\sigma_{b,i}^{\mathtt{mint}}$ for each $H\left(\kappa_{b,i}\right)$ are genuine;*
- *$V$ verifies that $H$ applied to each element of $\{\kappa_k\}_k$ is a subset of $\left\{H\left(\kappa_{b,i}\right)\right\}_{b,i}$;*
  - *If it is $\zeta - |\mathcal{K}| < \xi$ the banknote is considered* used up *and the verification fails;*
  - *Otherwise, $V$ chooses a* random *subset $\mathcal{L} \subseteq \mathcal{J}$ such that $|\mathcal{L}| = \xi$. Then, for each $\ell \in \mathcal{L}$, let $b_\ell$ be a randomly chosen bit.*
    - *For each $\ell \in \mathcal{L}$, compute $\mathtt{OTM}_\ell(b_\ell)$ by measuring the corresponding $|\mathtt{OTM}\rangle_\ell$;*
    - *$V$ then verifies that $H(\mathtt{OTM}_\ell(b_\ell)) = H\left(\kappa_{b_\ell,\ell}\right)$. If this is not the case, the verification fails.*
  - *If the verification passes for each $\ell$, $V$ sets $\mathcal{K}' = \mathcal{K} \cup \mathcal{L}$ and $\mathcal{J}' = \mathcal{J} \setminus \mathcal{L}$: i.e. the opened OTMs are removed from the banknote and their pre-images are added to the set of known pre-images;*
  - *$V$ returns the banknote $\left(\left\{H\left(\kappa_{b,i}\right), \sigma_{b,i}^{\mathtt{mint}}\right\}_{b,i}, \mathcal{J}', \left\{\mathtt{OTM}_j, |\mathtt{OTM}\rangle_j\right\}_{j \in \mathcal{J}'}, \zeta, \{\kappa_k\}_{k \in \mathcal{K}'}\right)$.*

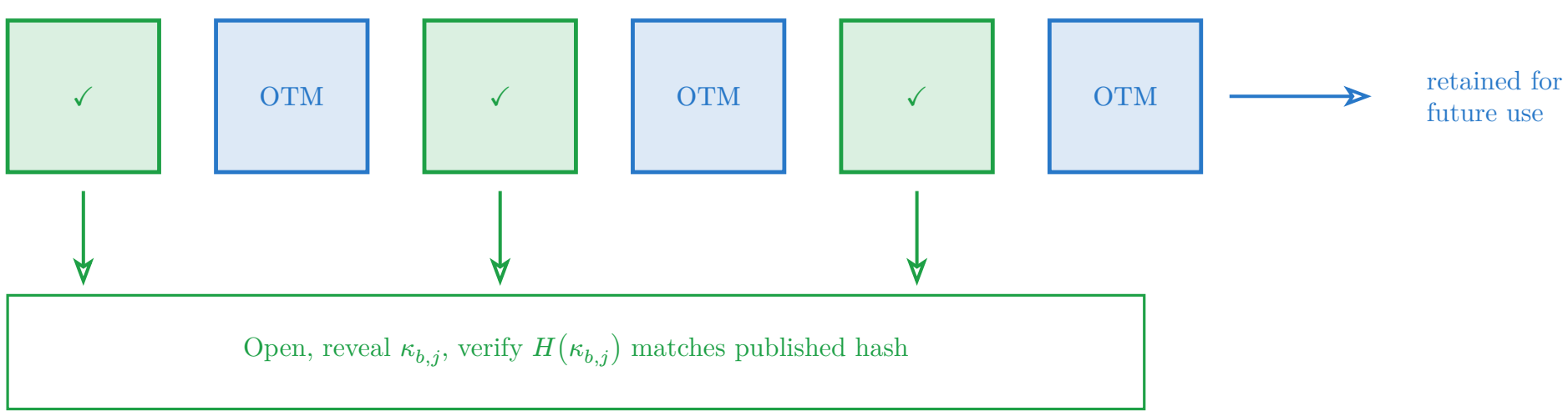

**Fig. 3.** Cut-and-choose verification. Green: OTMs randomly selected, opened, and checked. Blue: OTMs retained for future verifications.

**Theorem 1.** *(Unforgeability) Let $H$ be a preimage-resistant hash function and assume the OTM scheme is simulation-secure. Then, for any QPT adversary $\mathcal{A}$ given a single valid quantum banknote, the probability that $\mathcal{A}$ outputs two banknotes that both pass verification is negligible in $\xi$.*

*Proof.* Let $\mathcal{A}$ be a QPT adversary that receives a single valid quantum banknote with $|\mathcal{J}|$ unused OTMs and outputs two candidate banknotes $B_1, B_2$. We show that the probability both pass verification is negligible by a reduction to preimage resistance of $H$ and simulation-security of the OTM scheme.

**Step 1: Partition of OTM states via simulation-security.** Each OTM $\mathtt{OTM}_j$ for $j \in \mathcal{J}$ is backed by a quantum state $|\mathtt{OTM}\rangle_j$. By simulation-security (see Appendix A), any QPT adversary interacting with $\mathtt{OTM}_j$ can be simulated by a simulator that makes *at most one query* to the ideal functionality $\mathcal{F}_{\mathrm{OTM}}$. In particular, no adversarial strategy — including entangling the OTM state across $B_1$ and $B_2$ — can extract information about both $\kappa_{0,j}$ and $\kappa_{1,j}$ from a single $\mathtt{OTM}_j$, except with negligible probability. It follows that, up to negligible error, each OTM is *effectively used by at most one* of $B_1, B_2$. We may therefore define $S_1 \subseteq \mathcal{J}$ to be the set of indices whose OTM is effectively assigned to $B_1$ (i.e., $B_1$'s strategy for index $j$ is simulable using a query to $\mathcal{F}_{\mathrm{OTM}}$), and $S_2 = \mathcal{J} \setminus S_1$ those assigned to $B_2$, so that $S_1 \cup S_2 = \mathcal{J}$ and $S_1 \cap S_2 = \emptyset$.

**Step 2: Reduction to preimage resistance.** Note that the partition $(S_1, S_2)$ is determined at the moment $\mathcal{A}$ produces $B_1$ and $B_2$, since the quantum states must be physically embedded in the banknotes before they are handed to the verifiers. In particular, $(S_1, S_2)$ is fixed *before* the verifier challenge sets $\mathcal{L}_1, \mathcal{L}_2$ are drawn, regardless of whether the two verifications occur simultaneously or sequentially.

Consider the verification of $B_1$. The verifier selects a uniformly random subset $\mathcal{L}_1 \subseteq \mathcal{J}$ with $|\mathcal{L}_1| = \xi$ and, for each $j \in \mathcal{L}_1$, a uniformly random challenge bit $b_j \in \{0,1\}$. For $j \in \mathcal{L}_1 \cap S_1$, banknote $B_1$ holds the real OTM state and can produce a valid pre-image. For $j \in \mathcal{L}_1 \setminus S_1$, banknote $B_1$ does *not* hold the real quantum state for $\texttt{OTM}_j$. To pass the verification check $H\big(\texttt{OTM}_j(b_j)\big) = H\big(\kappa_{b_j,j}\big)$, $\mathcal{A}$ must produce some $\kappa'$ with $H(\kappa') = H\big(\kappa_{b_j,j}\big)$. Since $|\kappa_{b,i}| - |H\big(\kappa_{b,i}\big)| = \omega(\log \lambda)$ (see Construction 4.4), the pre-images have superlogarithmic min-entropy conditioned on their hash values, and so guessing a valid pre-image succeeds with negligible probability. Producing such $\kappa'$ without access to $\texttt{OTM}_j$ therefore constitutes a break of the preimage resistance of $H$. We construct the reduction $\mathcal{B}$ as follows: given a preimage-resistance challenge $y^*$, $\mathcal{B}$ embeds $y^*$ into a uniformly random position of the banknote's hash table (replacing one $H\big(\kappa_{b_j,j}\big)$), simulates the remaining OTMs honestly, runs $\mathcal{A}$, and extracts a pre-image from $\mathcal{A}$'s output for the embedded position whenever $\mathcal{L}_1 \nsubseteq S_1$. The reduction succeeds with probability at least $1/(2\zeta)$ times $\mathcal{A}$'s advantage (the factor accounts for the random choice of which position and bit to embed).

An identical argument applies to $B_2$ with respect to $\mathcal{L}_2$ and $S_2$.

**Step 3: Probability bound.** Absent a break in preimage resistance, $\mathcal{A}$ succeeds only if $\mathcal{L}_1 \subseteq S_1$ *and* $\mathcal{L}_2 \subseteq S_2$. Since $\mathcal{L}_1$ and $\mathcal{L}_2$ are chosen uniformly and independently at random, and the partition $(S_1, S_2)$ is fixed before the challenge sets are drawn, we have:

$$\Pr[\mathcal{L}_1 \subseteq S_1 \wedge \mathcal{L}_2 \subseteq S_2] = \frac{\binom{|S_1|}{\xi}}{\binom{|\mathcal{J}|}{\xi}} \cdot \frac{\binom{|S_2|}{\xi}}{\binom{|\mathcal{J}|}{\xi}}.$$

Subject to $|S_1| + |S_2| = |\mathcal{J}|$, this product is maximized when $|S_1| = |S_2| = |\mathcal{J}|/2$ (by the log-concavity of binomial coefficients). To see the bound, note that each ratio satisfies $\binom{n/2}{\xi}/\binom{n}{\xi} = \prod_{i=0}^{\xi-1}(n/2 - i)/(n-i) \le (1/2)^{\xi}$, since each factor $(n/2 - i)/(n-i) \le 1/2$ for $0 \le i < n/2$. Hence:

$$\frac{\binom{|\mathcal{J}|/2}{\xi}^2}{\binom{|\mathcal{J}|}{\xi}^2} \le \left(2^{-\xi}\right)^2 = 2^{-2\xi},$$

which is negligible for $\xi = \omega(\log \lambda)$.

**Conclusion.** Combining Steps 2 and 3, the probability that $\mathcal{A}$ succeeds is at most $2^{-2\xi} + \nu(\lambda)$, where $\nu(\lambda)$ is a negligible function arising from the preimage resistance of $H$ and the simulation-security of the OTM scheme. For $\xi = \omega(\log \lambda)$, this is negligible in the security parameter.

Notice that $V$ always needs to verify that the still unused OTMs aren't tampered with, as this is what ensures the banknote is not double-spent. With regard to this, the parameter $\mathcal{J}$ expresses how used up a banknote is. The mint could implement a policy where a banknote can be returned when $\mathcal{J}$ is under a certain threshold.

**On public verifiability.** We emphasize that "publicly verifiable" means any party — not just the mint — can verify a banknote, in contrast to Wiesner's private-key scheme. Verification is, however, *consumptive*: each verification uses up a portion of the banknote's OTMs. This is inherent to our cut-and-choose approach, not a limitation that can be removed without fundamentally changing the construction.

**On limited verifications.** The bounded number of verifications is a fundamental property of our scheme, not merely a parameter choice. Each verification consumes $\xi$ OTMs, limiting the banknote's lifetime to roughly $\frac{\zeta}{\xi}$ verifications. In practice, the mint would set $\zeta$ based on expected transaction frequency. A "spent" banknote (where $|\mathcal{J}| < \xi$) must be returned to the mint for *redemption*: the holder provides the remaining classical data, and the mint issues a fresh banknote. This redemption step is analogous to exchanging worn physical currency and can be implemented via the same channels used for initial minting.

The quantum banknotes can be readily transferred from the mint to some other party $P$, and more in general between any two parties $P_s, P_r$.

**Construction 4.6.** *(Transferring Quantum Money) To transfer a quantum banknote* $\left(\left\{H(\kappa_{b,i}), \sigma_{b,i}^{\mathtt{mint}}\right\}_{b,i}, \mathcal{J}, \left\{\mathtt{OTM}_j, |\mathtt{OTM}\rangle_j\right\}_j, \zeta, \{\kappa_k\}_k\right)$ *from* $P_s$ *to* $P_r$*,* $P_s$ *copies all the classical data* $\left(\left\{H(\kappa_{b,i}), \sigma_{b,i}^{\mathtt{mint}}\right\}_{b,i}, \mathcal{J}, \zeta, \{\kappa_k\}_k\right)$ *and sends it to* $P_r$*.* $P_s$ *also transfers the one-time memories* $\left\{\mathtt{OTM}_j, |\mathtt{OTM}\rangle_j\right\}_j$ *to* $P_r$*. Notice that* $P_s$ *has lost the ability to double spend the banknotes because the states* $\left\{|\mathtt{OTM}\rangle_j\right\}_j$ *cannot be cloned.*

As shown in Theorem 1, the security of the scheme is a direct consequence of (1) the simulation-security of the one-time memories and (2) the preimage resistance of the hash function $H$.

## 5 Upgrading to Quantum Tokens for Digital Signatures

As noted by Ben-David and Sattath [36], public key quantum money schemes based on subspace states can be turned into *quantum tokens for digital signatures (QTDS)* with minimal modifications. A quantum token for digital signatures is a quantum state that allows its holder to sign a single message of their choice *on behalf of the issuer*. Anyone can verify the validity of the signature using only classical resources. Moreover, the token can be *verified* for its ability to sign a message, without destroying its signing power.

Notice that, in comparison to more powerful schemes such as *quantum one-shot signatures* [37, 38], quantum tokens do not prevent the minting party from issuing multiple tokens. This makes quantum tokens unsuitable for some applications: for voter delegation in consensus algorithms, validators must trust that no voting token has been issued multiple times. Nevertheless, quantum tokens are useful when we **do want** the mint to issue multiple tokens, such as in quantum money where different signatures can distinguish between banknote series and denominations.

Among the possible applications, one that stands out is *money commitments*, where a party uses the signature mechanism in the banknote to sign a given document, thereby committing their money to it without using a blockchain. Consider a notary setting: Alice wishes to prove she held funds at a specific time.

She signs a timestamp and document hash using her quantum token; this signature is verifiable by anyone with the mint's public key, yet the act of signing consumes the token's signing power, preventing reuse. For accounting, a company could sign revenue receipts, creating an auditable trail of fund allocations.

Betting provides another compelling application. Here, the mint acts as a trusted casino or betting platform. A bettor receives quantum tokens representing their stake. To place a bet, they sign their prediction (e.g., "Team A wins"). The casino collects these signatures; winners receive fresh tokens, while losing tokens are already spent. The quantum nature prevents double-betting: once you sign a prediction, that token cannot sign the opposite outcome.

To show that our quantum money scheme can be turned into a QTDS for signing a single bit, we require minimal modifications. To sign for multiple bits, one can simply run multiple instances of the single-bit signing protocol in parallel.

**Construction 5.7.** *(Upgrading to Quantum Tokens for Digital Signatures) The minting algorithm in Construction 4.4 and transfer algorithm in Construction 4.6 remain unchanged. The verification algorithm in Construction 4.5 remains mostly unchanged except that, now, we require* $|\mathcal{J}| > \frac{\zeta}{2} + 1$ *to ensure that a majority of unopened OTMs remain after verification.*

*Then, signing works as follows for quantum banknote* $\left(\left\{H\left(\kappa_{b,i}\right), \sigma^{\mathtt{mint}}_{b,i}\right\}_{b,i}, \mathcal{J}, \left\{\mathtt{OTM}_j, |\mathtt{OTM}\rangle_j\right\}_j, \zeta, \{\kappa_k\}_k\right)$ *and bit* $\beta \in \{0,1\}$*:*

- *For each* $i \in \mathcal{J}$*, compute* $\mathtt{OTM}_i(\beta)$ *by measuring the corresponding* $|\mathtt{OTM}\rangle_i$*;*
- *The signature on message* $\beta$ *is then the tuple* $\left(\beta, \mathcal{J}, \{\mathtt{OTM}_i(\beta)\}_{i\in\mathcal{J}}, \left\{H\left(\kappa_{b,i}\right), \sigma^{\mathtt{mint}}_{b,i}\right\}\right)$.

The verification of a signature works as follows:

**Construction 5.8.** *(Verifying QTDS Signatures) The verification algorithm for signature* $\left(\beta, \mathcal{J}, \{\mathtt{OTM}_i(\beta)\}_{i\in\mathcal{J}}, \left\{H\left(\kappa_{b,i}\right), \sigma^{\mathtt{mint}}_{b,i}\right\}\right)$ *is as follows:*

- *Check that* $|\mathcal{J}| > \zeta/2 + 1$*;*
- *Verify that the signatures* $\sigma^{\mathtt{mint}}_{b,i}$ *for each* $H\left(\kappa_{b,i}\right)$ *are genuine;*
- *Verify that for strictly more than* $|\mathcal{J}|\ /2$ *indices* $i \in \mathcal{J}$*, the hash* $H(\mathtt{OTM}_i(\beta))$ *is contained in* $\left\{H\left(\kappa_{b,i}\right)\right\}_{b,i}$.

**Corollary 1.** *(QTDS Unforgeability) Let* $H$ *be a preimage-resistant hash function and assume the OTM scheme is simulation-secure (see Appendix A). For any QPT adversary* $\mathcal{A}$ *that receives a quantum token and signs a bit* $\beta \in \{0,1\}$*, the probability that* $\mathcal{A}$ *also produces a valid signature on* $1-\beta$ *is negligible.*

*Proof.* Suppose $\mathcal{A}$ signs bit $\beta$, producing valid pre-images $\left\{\kappa_{\beta,i}\right\}_{i\in\mathcal{J}}$ for a majority of indices. To forge a signature on $1-\beta$, $\mathcal{A}$ must produce valid pre-images $\left\{\kappa_{1-\beta,i}\right\}_{i\in\mathcal{J}'}$ for some $\mathcal{J}'$ with $|\mathcal{J}'| > |\mathcal{J}|\ /2$.

By simulation-security of the OTM scheme (see Appendix A), each $\mathtt{OTM}_i$ can be simulated by a single query to $\mathcal{F}_{\mathrm{OTM}}$. Since $\mathcal{A}$ already queried each OTM on bit $\beta$ to produce the first signature, the simulator has consumed each OTM's single query. Therefore, $\mathcal{A}$ cannot extract $\kappa_{1-\beta,i}$ from $\mathtt{OTM}_i$ for any $i$, except with negligible probability.

To produce $\kappa'$ with $H(\kappa') = H\left(\kappa_{1-\beta,i}\right)$ without OTM access, $\mathcal{A}$ must break the preimage resistance of $H$. Since $|\kappa_{b,i}| - \left|H\left(\kappa_{b,i}\right)\right| = \omega(\log\lambda)$, guessing succeeds with negligible probability. By a union bound over at most $\zeta$ indices, the total forging probability is negligible.

## 6 Bottlenecks and Future Work

**Pre-transfer verification.** Perhaps the most significant open problem is enabling *pre-transfer verification*: allowing a receiver to verify banknote authenticity *before* accepting the quantum states from a potentially malicious sender. Currently, the revealed pre-images $\{\kappa_k\}_k$ in Construction 4.5 are classical, allowing inspection before transfer, but this only provides assurance when the sender is honest. Solving this would require the receiver to somehow verify the quantum states without taking possession, likely requiring additional rounds of interaction or cryptographic commitments.

**Resource scaling.** Our scheme's quantum resource consumption grows linearly with the maximum number of verifications $N$. Reducing this to sub-linear growth, while maintaining minimal quantum computational requirements, is an interesting open question.

**Infinite verifiability.** A strengthening of the previous point, where the scheme can be verified an arbitrary number of times.

**Weaker hardware assumptions.** Our reliance on trusted hardware for OTM instantiation is a limitation. Furthermore, OTMs can also be implemented on trusted hardware without making use of conjugate coding or any other kind of quantum resource. The main difference is that, in the latter case, forgery can be performed at any time: for instance, a malicious actor could perform a memory dump [39] and extract the relevant information to clone a banknote at a later time. On the contrary, relying on quantum resources makes it so that an attack can be performed only *before* the states are actually measured, as specific measurements are needed to extract enough information from the qubits to reconstruct the OTMs. In any case, developing OTM constructions with weaker assumptions, or replacing hardware trust with limited quantum computational assumptions, would broaden applicability.

**Applications.** Finally, we are interested in exploring additional applications of quantum tokens and in experimental implementations of our protocols.

## A Appendix: Simulation-Secure OTM Definition

We recall the ideal functionality for one-time memories, originally introduced by Goldwasser, Kalai, and Rothblum [16]. The quantum UC-secure construction is due to Broadbent, Gharibian, and Zhou [22].

**Functionality A.9.** *($\mathcal{F}_{\mathrm{OTM}}$)* ***Create:*** *Upon input $(s_0, s_1)$ from the sender, with $s_0, s_1 \in \{0,1\}$, send* `create` *to the receiver and store $(s_0, s_1)$.*

***Execute:*** *Upon input* $b \in \{0,1\}$ *from the receiver, send* $s_b$ *to the receiver. Delete any trace of this instance.*

A protocol $\Pi$ *UC-realizes* $\mathcal{F}_{\mathrm{OTM}}$ if there exists a simulator $\mathcal{S}$ such that for any quantum environment $\mathcal{Z}$, the real execution of $\Pi$ is statistically indistinguishable from the ideal execution with $\mathcal{F}_{\mathrm{OTM}}$ and $\mathcal{S}$.

Informally, simulation security guarantees that a malicious receiver learns at most one of $s_0, s_1$, since any attack can be simulated by making a single query to the ideal functionality. The functionality above is stated for single-bit secrets; it extends to bitstring secrets $s_0, s_1 \in \{0,1\}^n$ by parallel composition of $n$ independent instances.